\documentclass[aps,prx,twocolumn,superscriptaddress,longbibliography]{revtex4-2}
\usepackage{float,graphicx,amssymb,amsfonts,amsmath,chemarr,color,commath,braket,hyperref,bm,bbold,dsfont}
\usepackage[normalem]{ulem}
\allowdisplaybreaks[1]

\newcommand{\beginsupplement}{%
\setcounter{page}{1}
        \setcounter{equation}{0}
        \renewcommand{\theequation}{S\arabic{equation}}%
        \setcounter{table}{0}
        \renewcommand{\thetable}{S\arabic{table}}%
        \setcounter{figure}{0}
        \renewcommand{\thefigure}{S\arabic{figure}}%
     }

\begin{document}

\title{Spatial structure promotes coexistence, resilience, and synergy in microbial communities}

\author{Motasem ElGamel}
\email{motasem.elgamel@emory.edu}
\affiliation{Department of Physics and Initiative in Theory and Modeling of Living Systems, Emory University, Atlanta, Georgia 30322, USA}
\affiliation{Department of Physics and Astronomy, University of Pittsburgh, Pittsburgh, Pennsylvania 15260, USA}

\author{Andrew Mugler}
\affiliation{Department of Physics and Astronomy, University of Pittsburgh, Pittsburgh, Pennsylvania 15260, USA}

\begin{abstract}
Microbial communities are highly complex systems with a tremendous number of interactions. They thrive in vastly different environments, from the human body to soil and marine ecosystems, and show remarkable properties such as coexistence, diversity, and resilience. Additionally, they are almost always spatially structured. However, the role of space in shaping microbial communities remains poorly understood, in part because explicitly modeling space substantially increases model complexity and limits further exploration of its effects across different settings. Here, we present a minimal theoretical model that incorporates spatial structure into a consumer-resource framework through local and global nutrient consumption. Using this model, we explore the effect of spatial structure on several important properties of microbial communities. We find that spatial structure promotes coexistence under nutrient limitation, increases community resilience to environmental fluctuations, and allows communities to benefit synergistically from poor nutrients. We then extend the model to make spatial structure adaptive, allowing species to become more isolated when competition threatens their presence. Our results show that moderate spatial adaptation promotes coexistence, while weak and strong adaptation do not. Interestingly, strong spatial adaptation can substantially increase community coexistence under environmental fluctuations, while not necessarily promoting coexistence in the absence of fluctuations. Together, our results provide a simple framework for exploring how spatial structure shapes microbial community assembly, dynamics, and response to fluctuating environments.
\end{abstract}

\maketitle

\section{introduction}

Microbes are among the most abundant forms of life on Earth \cite{bar2018biomass}. They form highly complex communities with many interacting components on all scales. Such complexity gives rise to many phenomena of interest and has inspired many questions in both experimental and theoretical research, particularly in ecology \cite{widder2016challenges, van2022ecological}. In nature, microbial communities are almost always spatially structured \cite{warrier2026interplay}. Spatial structure has been observed in many distinct microbial systems. For example, it has been observed in the gut microbiomes of different species \cite{schlomann2021gut,henderson2025disentangling}, as well as in soil microbial communities \cite{raynaud2014spatial} and the human oral microbiome \cite{shi2020highly,mark2016biogeography}. The biological processes responsible for the emergence of spatial structure differ across systems and environments \cite{warrier2026interplay,henderson2025disentangling,yanni2019drivers}. Regardless of its origin, spatial structure is observed consistently across distinct environments and is considered an important property of microbial communities \cite{warrier2026interplay,backer2026spatial}. Indeed, this spatial organization influences microbial function, interactions among species and with their environment, abundance, evolution, competition, stability, and coexistence \cite{warrier2026interplay,momeni2013strong,backer2026spatial,pickett1995landscape,mattei2024exploring}.

Significant progress has been made both experimentally and theoretically in understanding the effects of space on microbial coexistence and dynamics \cite{lobanov2023spatial,denk2025spatial,weiner2019spatial,kim2008defined}. Theoretically, generalized Lotka-Volterra (LV) \cite{cui2024houches,bunin2017ecological} and consumer-resource (CR) models \cite{cui2024houches,chesson1990macarthur} have been adapted to account for space. Previous work has relied on explicit modeling of space, either as patches \cite{al2026spatiotemporal,garcia2024interactions} or as non-overlapping spatial territories between which nutrients can diffuse \cite{weiner2019spatial}. While elegant, such complex frameworks can limit the utility of these models in exploring wider implications of spatial structure. Hence, a simpler and more abstract model for spatial structure is needed to capture the essential properties of spatially structured microbial communities while allowing further exploration of its consequences.

\begin{figure}
\includegraphics[width=1\columnwidth]{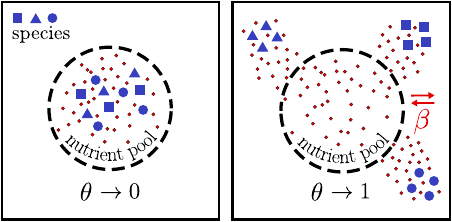}
\caption{(left) The traditional CR model is recovered in the well-mixed limit of spatial CR model. (right) The full separation limit of the model where species are in local spatial niches. The parameter $\beta=D/D_0$ determines the rate of nutrient exchange between the global nutrient pool and the local niches.}
\label{mdl}
\end{figure}

\begin{figure*}
\includegraphics[width=1\linewidth]{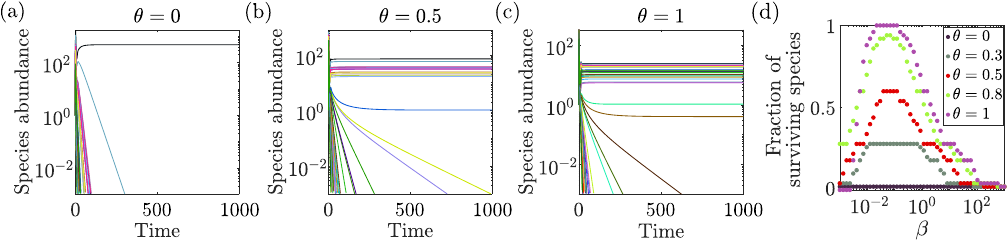}
\caption{(a)-(c) Species abundance dynamics at $\theta=0$, $\theta=0.5$, and $\theta=1$. For each simulation, all species had the same value of $\theta_i=\theta$. Each color represents a different species. (d) Fraction of surviving species vs. the nutrient exchange rate between the global pool and local niches rescaled by the rate of external nutrient supply, $\beta$, shown for different levels of community separation, $\theta$. All curves and data points were obtained by numerically solving the model \cite{data}. See \cite{paramchoice} for parameter choices.}
\label{coexist}
\end{figure*}

One important effect of spatial structure is that it changes the scale over which nutrients are shared and, consequently, the scale of species competition \cite{kim2008defined, dal2020short}. When nutrients are consumed locally much faster than they diffuse across the community, species interact through the resources available within their local spatial niches. In contrast, when nutrients diffuse rapidly compared to consumption, they become effectively global and are shared among species, leading to more global competition. Therefore, at moderate rates of nutrient diffusion in the community, species interact more locally in spatially structured communities and globally in well-mixed communities. This provides a possible coarse-grained description of space based on the relative contributions of local and global nutrient consumption.

Here, we present a simple model in which space is not explicit, but its effect is accounted for through one parameter. Building on MacArthur's CR model \cite{chesson1990macarthur}, in which species interactions are mediated through their nutrient preferences and consumption, our model accounts for space through two possible modes of consumption, global and local, as well as intermediate combinations of the two [Fig.\ \ref{mdl}].

\section{model}

\begin{figure*}
\centering
\includegraphics[width=1\linewidth]{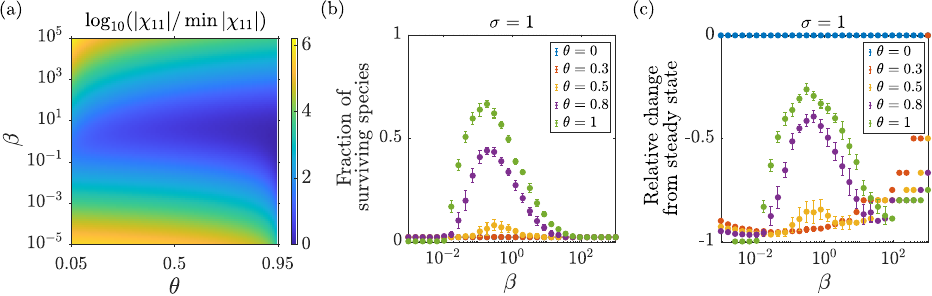}
\caption{(a) Heatmap of susceptibility obtained for a simple model of two species with equal nutrient preferences and mortality rates. It shows the susceptibility of the first species due to perturbations in its own mortality rate, $\chi_{11}=\delta N^*_1/\delta k_1$, at different $\beta$ and $\theta$ values normalized by its minimum. Species preferences were chosen as $\alpha_1=\alpha_2=5$. (b) The fraction of surviving species at stationary state in the presence of environmental noise. Maximum survival is obtained at $\theta=1$. (c) The relative change in the number of surviving species after introducing noise, defined as the difference between the number of surviving species in the stationary state after adding noise and the number of surviving species at the deterministic steady state before adding noise. In panels (b) and (c), data points are the average of 20 stochastic trajectories with the same nutrient preferences and error bars are the standard deviation \cite{data}. See \cite{paramchoice} for parameter choices.}
\label{susc}
\end{figure*}

A simple form of the traditional CR model for a community of $S$ species and one resource, $R$, is given by
\begin{align}
    \frac{dN_i}{dt}&=(r_i R -m_i) N_i, \label{CRmdlN}\\
    \frac{dR}{dt}&=-\gamma R \sum_{i=1}^S r_i N_i + D_0 (R_0 - R), \label{CRmdlR}
\end{align}
where $N_i$ is the abundance of species $i$ that grows by consuming $R$ with a preference, $r_i$. The environment sets the mortality rate, $m_i$, and $1/\gamma$ sets the conversion factor between nutrients and abundances. Here, we assume that the resource is externally supplied to the environment at a rate $D_0$, and $R_0$ is the supplied nutrient concentration or the background nutrient concentration in the absence of consumers. $S$ is the number of species in the community. This model gives the canonical result of the competitive exclusion principle, in which only one species will survive in the presence of only one nutrient \cite{hardin1960competitive}. In the rest of our analysis, we adopt the CR model with an externally supplied resource, which describes abiotic resources that cannot self-replicate \cite{cui2020effect}. We expect all results to hold under different nutrient supply dynamics.
 
Next, we extend the model in Eqs.\ (\ref{CRmdlN}) and\ (\ref{CRmdlR}) to include spatial structure in the community. We assume that the community is split into local niches, and the global resource pool, $R$, is allowed to exchange nutrients with these local niches at a rate $D$. The concentration of $R$ available in each local niche is denoted by $T_i$. In this sense, $R$ is a global nutrient and $T_i$ are local nutrients available to species in their local niches [Fig.\ \ref{mdl}]. In the rest of the paper, we rescale all rates by $D_0$ (see supplementary material for unscaled model \cite{supp}), and define $\tau=D_0t$, $\alpha_i=r_i/D_0$, $k_i=m_i/D_0$, and $\beta=D/D_0$. The updated spatial CR model after rescaling is then given by
\begin{align}
    \frac{dN_i}{d\tau}&=[(1-\theta_i)\alpha_i R + \theta_i \alpha_i T_i -k_i] N_i, \label{spatialCRmdlN}\\
    \frac{dR}{d\tau}&=-\gamma R \sum_{i=1}^S (1-\theta_i) \alpha_i N_i -\beta \sum_{i=1}^S (R-T_i) \nonumber\\&+ (R_0 - R), \label{spatialCRmdlR}\\
    \frac{dT_i}{d\tau}&=\beta (R-T_i) -\gamma \theta_i \alpha_i T_i N_i, \label{spatialCRmdlT}
\end{align}
where $0 \le \theta_i \le 1$ determines each species' preference for the local nutrient. It can also be interpreted as the fraction of species $i$ that consumes from its local spatial niche, i.e. the fraction that is isolated from global nutrient competition. In the limit $\theta_i \rightarrow 0$, we recover the well-mixed model in Eqs.\ (\ref{CRmdlN}) and\ (\ref{CRmdlR}), and in the opposite limit, $\theta_i \rightarrow 1$, the community is fully spatially separated [Fig.\ \ref{mdl}].

Moving forward, we fully exploit the simplicity of this model and use it as a framework to investigate the effect of spatial structure on different aspects of microbial community assembly, dynamics, and interactions with the environment. In particular, we focus on the effects of spatial structure on microbial coexistence under limited nutrients and resilience to environmental fluctuations. We then further extend the model to account for adaptation of community spatial structure.

\section{results}

\subsection{Coexistence}

As the degree of spatial separation increases with $\theta$, global competition weakens and local competition is strengthened. As a result, coexistence becomes possible and the community can support more than the one species implied by the competitive exclusion principle. Figs. \ \ref{coexist}(a)-(c) show the species abundance dynamics at different values of $\theta$ when $\theta_i=\theta$ is the same for all species. Coexistence peaks at full spatial separation ($\theta=1$), when each species consumes nutrients only from its local niche.

Additionally, coexistence depends on how quickly nutrients diffuse into local niches compared to the rate at which they are supplied from the external environment, as given by $\beta$ [Fig.\ \ref{coexist}(d)]. The fraction of surviving species, defined as the final number of surviving species at steady state divided by the initial number of species, peaks at intermediate values of $\beta$ across different values of $\theta$. This means that the effect of spatial separation on coexistence depends on the environment, particularly on how permissive it is to nutrient diffusion. It has been previously proposed that habitat connectivity in soil has a direct effect on diversity, with poorly connected habitats promoting greater diversity than well-connected ones, particularly because diffusion-limited environments sustain nutrient gradients and shelter less competitive species \cite{long2005aquatic}.

Importantly, coexistence does not require complete spatial separation [Fig.\ \ref{coexist}(b) and (d)]. It can also arise at intermediate degrees of separation, where part of the population consumes locally while the rest competes for the global nutrient pool. In the simple case of a community of only two species, we can derive the condition for coexistence, defined as a stable steady state with positive abundance for both species simultaneously, in the limiting cases $\theta=0$ and $\theta=1$ as a function of model parameters \cite{supp}. In contrast to well-mixed communities ($\theta=0$), where a positive steady state abundance for one species necessarily gives a negative steady state abundance for the other, spatially separated communities ($\theta=1$) satisfy the coexistence condition \cite{supp}.

\begin{figure*}
\includegraphics[width=1\linewidth]{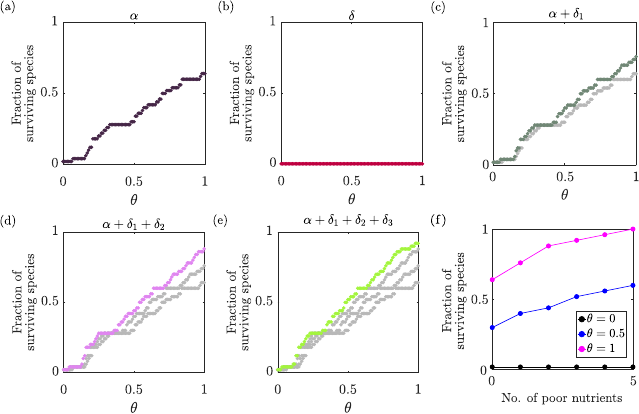}
\caption{Fraction of surviving species as a function of $\theta$ in the presence of only one nutrient for which all species have (a) a strong (rich nutrient) (b) or weak (poor nutrient) preference.  No species survive in the presence of only the poor nutrient. (c)-(f) Sequentially adding poor nutrients to the environment increases the number of surviving species in the presence of the rich nutrient compared to the original fraction of surviving species in the presence of only the rich nutrient. In the well-mixed limit ($\theta=0$), the presence of both nutrients does not increase survival, this synergistic effect only appears in the presence of spatial structure. Preferences for all poor nutrients were set to $\delta=0.1$ for all species. Data points were obtained by numerically solving the model \cite{data}. In all panels, other model parameters follow \cite{paramchoice}.}
\label{hetenv}
\end{figure*}

\subsection{Susceptibility}

Microbial communities in natural environments are consistently subject to perturbations and recurrent changes. Next, we use our model to explore the effect of spatial separation on community resilience to perturbations.

When a parameter is perturbed, the steady state of the community will change. For example, a perturbation in $k_j$ such that $k_j\rightarrow k_j+\delta k_j$, will result in a corresponding shift in the steady state species abundances, $N^*_i\rightarrow N^*_i + \delta N^*_i$ \cite{goyal2025universal}. We follow the analysis of Ref.\ \cite{goyal2025universal} and define the abundance susceptibility as $\chi_{ij}= \delta N^*_i/\delta k_j$. In other words, the susceptibility measures how sensitive species abundances are to perturbations in a given parameter. Parameter perturbations can reflect intrinsic (e.g. nutrient preference, fitness, etc) or extrinsic (environmental) effects. We choose the susceptibility to perturbations in the rescaled mortality rate $k_j$ to define species susceptibility to external perturbations because $k_j$ is set by a combination of species intrinsic and extrinsic environmental factors and sets the minimum rate required for population maintenance.

We use a simplified version of the model with only two species competing for one nutrient. This simple version allows us to analytically solve for susceptibility; for a detailed calculation see the supplemental material \cite{supp}. The result is illustrated in Fig.\ \ref{susc}(a), with the assumption that both species have exactly equal preferences for the nutrient. For unequal preferences, the qualitative result does not change as long as neither species is extinct.

In Fig.\ \ref{susc}(a) the species are most susceptible when $\beta \gg 1$, due to strong competition between species in this limit which makes a small change in the environment strongly affect abundances, and when $\beta \ll 1$, due to poor nutrient diffusion in this limit which results in small abundances that are naturally more susceptible to small changes. In the limit $\theta\rightarrow 1$, susceptibility significantly decreases at intermediate $\beta$ values due to the shift in competition for nutrients between species from global to local, compared to the well-mixed limit $\theta \rightarrow 0$. Also, note that the points at exactly $\theta=0$ and $\theta=1$ were excluded from the figure for visibility and to omit unphysical results. In particular, when $\theta=0$, susceptibility diverges regardless of the value of $\beta$, and when $\theta=1$ and $\beta\ll1$, susceptibility approaches zero, not due to a true decrease in susceptibility but because species go extinct due to lack of nutrients in their local niches. 

\subsection{Community resilience}

Now that we have a measure of the community response to one-time external environmental perturbations, we move to a more realistic scenario for microbial communities: fluctuating environments.

Natural environments fluctuate. These fluctuations can be in the form of nutrient availability in the environment or other extrinsic factors that affect species growth and mortality rates, and can have drastic effects on community coexistence and diversity \cite{mancuso2021environmental}. Here, we consider environmental fluctuations as fluctuations in the mortality rates, i.e. $k_i\rightarrow k_i + \sigma \eta_i$, where $\sigma$ is the noise strength and $\eta_i$ is white (uncorrelated) Gaussian noise, defined as $\langle \eta_i(t) \eta_j(t^\prime) \rangle=\delta_{ij}\delta(t-t^\prime)$. This is equivalent to a multiplicative noise term in the abundance dynamics, i.e. a term of the form $\sigma N_i \eta_i$, previously explored in different ecological contexts and in microbial and viral systems as a model of extrinsic (environmental) fluctuations \cite{sayin2026bacterial, al2026spatiotemporal,xiao2026environmental,henderson2026fluctuating}. The modified abundance dynamics equation becomes 
\begin{equation}
    \frac{dN_i}{d\tau}=[(1-\theta_i)\alpha_i R + \theta_i \alpha_i T_i -k_i + \sigma \eta_i] N_i,
\label{multnoise}
\end{equation}
with the nutrient dynamics remaining the same as in Eqs.\ \ref{spatialCRmdlR} and\ \ref{spatialCRmdlT}. Because spatial structure decreases the community's susceptibility to perturbations, we expect this result to carry over in the presence of environmental fluctuations and spatially structured communities to be less susceptible to noise.

To quantify community resilience to environmental fluctuations, we first simulated the species dynamics at different $\theta$ and $\beta$ without noise until they reached steady state. We then added noise, as in Eq.\ \ref{multnoise}, and simulated the dynamics until a stable number of surviving species was reached (illustrated in supplementary figure\ \ref{Sfig1:noise} \cite{supp}). As a measure of resilience to fluctuations, we calculated the relative change in the number of surviving species after introducing noise, defined as the difference between the number of surviving species in the stationary state after adding noise and the number of surviving species at the deterministic steady state before adding noise, normalized by the steady state number before adding noise.

The fraction of surviving species in the stationary state in the presence of noise is illustrated in Fig.\ \ref{susc}(b), and the relative change is illustrated in Fig.\ \ref{susc}(c). We see that the relative change is least negative for intermediate $\beta$, showing that the community is most resilient to fluctuations at intermediate diffusion as expected from Fig.\ \ref{susc}(a). The relative change becomes less negative as $\theta\rightarrow 1$ until it peaks at full separation ($\theta=1$). It is important to note that the increasing trend prominent at $\beta\gg1$ is not due to increased resilience, but rather to the very low number of species present in the community before adding noise (supplementary figures \cite{supp}). In this case, the community is already near its final state, and noise does not cause significantly more loss [Fig.\ \ref{susc}(b)]. Similarly, at $\theta=0$, there is only one surviving species before and after adding noise.

Intuitively, by weakening global competition for nutrients, spatial structure decreases the effect of ``lucky events'' that may occur in some niches, allowing the species occupying those niches to overtake the community. Indeed, in previous work, we showed that spatial structure and multiplicative noise are both essential for explaining the statistical patterns of species abundances and the high population richness observed in an experiment where various, equally fit, barcoded subpopulations coexist \cite{sayin2026bacterial}.

Having considered how spatial structure affects competition under environmental fluctuations, we next ask how it changes species survival when multiple nutrients are available.

\begin{figure*}
\includegraphics[width=1\linewidth]{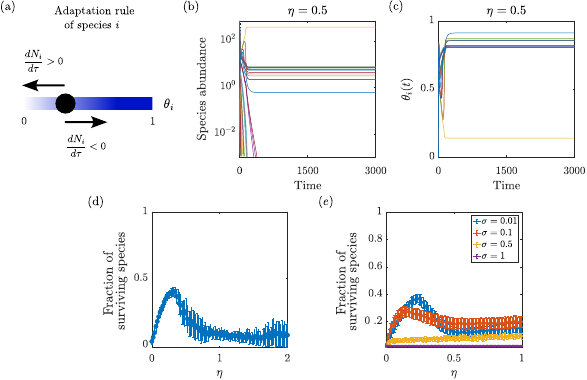}
\caption{(a) $\theta_i$ is allowed to vary for each species. Species adapt from global to local consumption (isolation) if they have a negative growth rate, and from local to global consumption if they have a positive growth rate. (b) Example of the abundance dynamics at an intermediate adaptation rate. Species were initialized at the same initial abundance of one. (c) The corresponding $\theta_i$ dynamics of the community simulated in (b) under the adaptation rule in (a). The community eventually settles into a spatial configuration where species are at different degrees of isolation and overlap in the community. (d) The fraction of surviving species peaks at an intermediate adaptation rate and is lower at both slow and fast adaptation rates (no noise, $\sigma=0$). Points show the mean across 20 independently sampled communities with different species-specific nutrient preferences. The same communities were used at every adaptation rate. Error bars represent the standard deviation. (e) Fraction of surviving species as a function of adaptation rate at different noise strengths. Very strong noise is harmful regardless of the adaptation rate. Moderately strong noise is beneficial at fast adaptation rates, but harmful at intermediate rates. See \cite{paramchoice} for simulation parameters.}
\label{adapt}
\end{figure*}

\subsection{Synergy with additional nutrients}

We start by extending the model to include more than one nutrient. The model becomes
\begin{align}
    \frac{dN_i}{d\tau}&=[(1-\theta_i) \sum_{j=1}^n \delta_{ij} R_j+\theta_i \sum_{j=1}^n \delta_{ij} T_{ij}   -k_i] N_i, \label{spatialmultinutrientN}\\
     \frac{dR_j}{d\tau}&=-\gamma R_j \sum_{i=1}^S (1-\theta_i) \delta_{ij} N_i -\beta \sum_{i=1}^S (R_j-T_{ij})\nonumber\\&+ (R_{j0} - R_j), \label{spatialmultinutrientR}\\
    \frac{dT_{ij}}{d\tau}&=\beta (R_j-T_{ij}) -\gamma \theta_i \delta_{ij} T_{ij} N_i, \label{spatialmultinutrientT}
\end{align}
where $R_j$ denotes nutrient $j$, $T_{ij}$ is the concentration of nutrient $j$ in niche $i$, $\delta_{ij}$ is the preference of species $i$ for nutrient $j$, $j$ is an index over nutrients, and $n$ is the total number of nutrients. It is assumed that all nutrients diffuse into the local niches at the same rate $\beta$. We consider only one resource for which all species have a strong preference (rich nutrient). All other added resources have weak preferences (poor nutrients). We refer to the species' preferences for the rich nutrient as $\alpha$, as used previously, while $\delta$ will refer to the species preferences for poor nutrients. A comparison between abundances in the presence of only the rich nutrient and only the poor nutrient is shown in Fig.\ \ref{hetenv}(a) and (b), respectively. The preference for the poor nutrient was fixed to a value such that $\delta_{ij}\ll k_i$ to intentionally prevent growth on the poor nutrient alone. Furthermore, we assumed that species have equal preferences for the poor nutrients, $\delta_{ij}=0.1$. We later relaxed this assumption and showed that the qualitative results of Fig.\ (\ref{hetenv}) remain unchanged (see supplementary figure\ \ref{Sfig2:hetenvpoor} \cite{supp}).

When both the rich and poor nutrients are present in the community, the result in the well-mixed limit ($\theta\rightarrow 0$) is the same as when only the rich nutrient is present. Strikingly, increasing $\theta$ increases species survival in the presence of a larger number of poor nutrients, until the number of surviving species is maximal at full separation [Figs.\ \ref{hetenv}(c)-(f), see \cite{supp} for detailed analysis]. Intuitively, the presence of isolated spatial niches shifts the problem from maximizing survival through competition, which would require finding the best nutrient strategy to outcompete other species present in the community, to dividing resources across local niches to maximize their benefit. In the latter case, even a small contribution from additional nutrients can allow a more generous division of resources across space and synergistically support more niches.

Next, we extend the model to allow for spatial adaptation, defined as the ability of the community to change its spatial configuration to promote survival.

\subsection{Adaptation}

Microbial communities are adaptive, and this adaptation is not just limited to changes in their metabolism or nutrient preferences, but can also involve changes in their spatial configuration, through, e.g., chemotaxis and migration \cite{keegstra2022ecological, warrier2026interplay}. These forms of motility can give rise to different spatial configurations. In the language of our model, we define different spatial configurations as the collection of species-specific values of $\theta_i$. To account for these spatial adaptation dynamics, we allow $\theta_i$ to vary dynamically and respond to species growth [Fig.\ \ref{adapt}(a)]. It was previously shown that competition avoidance (species favoring isolation when threatened) results in microbial spatial separation and favors coexistence \cite{mattei2026nonlinear,mattei2024exploring}. Similarly, we use competition avoidance as the basis for species adaptation in our model. The only exception is that we allow species that are not outcompeted by other species (i.e. able to maintain positive growth despite the competition) to favor no isolation. Hence, we define the dynamics of $\theta_i$ as
\begin{equation}
\frac{d\theta_i}{d\tau}=\begin{cases} 
	-\eta (1-\theta_i) \frac{1}{N_i}\frac{dN_i}{d\tau}, 	& \frac{dN_i}{d\tau} < 0,\\
      -\eta \theta_i \frac{1}{N_i}\frac{dN_i}{d\tau}, 	& \frac{dN_i}{d\tau}\geq 0,
   \end{cases}
\label{dyneta}
\end{equation}
where $\eta$ is the adaptation rate, and $(1/N_i)dN_i/d\tau$ is the instantaneous growth rate of species $i$. The prefactors $\theta_i$ and $1-\theta_i$ ensure that $\theta_i$ remains bounded between 0 and 1. When species have a positive growth rate and are outcompeting other species, the adaptation rule shifts their $\theta_i$ toward 0, increasing their access to the global nutrient pool and competition with other species [Fig.\ \ref{adapt}(a) and Eq.\ \ref{dyneta}]. Conversely, if species are losing the competition (i.e. their growth rate is negative), the adaptation rule shifts their $\theta_i$ toward 1, increasing their isolation from global competition and promoting survival [Fig.\ \ref{adapt}(a) and Eq.\ \ref{dyneta}]. Abundance and $\theta_i$ dynamics are illustrated in Figs.\ \ref{adapt}(b) and (c) at an intermediate adaptation rate $\eta=0.5$. Species were initialized with equal abundances, $N_i=1$, different preferences for the same nutrient \cite{paramchoice}, and $\theta_i=0$. As abundances change, the configuration of the community changes and different species become isolated to different degrees, allowing more species to survive.

Fig.\ \ref{adapt}(d) illustrates the fraction of surviving species at different adaptation rates. Counterintuitively, strong adaptation is not beneficial. This is because strong adaptation rapidly drives species toward either full isolation, $\theta_i=1$, or no isolation, $\theta_i=0$. As the community adapts toward a spatial configuration, the more competitive species (which have the strongest preference for the nutrient) can remain near $\theta_i=0$ and outcompete other species for the global nutrient pool. Because local niches only get nutrients from the global pool, a competitive species that is able to adapt very quickly and dominate the global pool will consume all available nutrients and competitive exclusion will apply. On the other hand, very weak adaptation does not allow species to adapt quickly enough before competitive exclusion occurs. Dynamics for the two extremes are illustrated in the supplementary figures \cite{supp} for $\eta=0.001$ and $\eta=1.5$. Intermediate adaptation allows less competitive species to grow just enough before being limited by isolation. This limits the expansion of the more competitive species forcing them towards more isolation (higher $\theta_i$) [Fig.\ \ref{adapt}(b) and (c)]. These results assume a well-mixed initial state ($\theta_i=0$) with equal abundance ($N_i=1$) for all species.

Next, we test the adaptation dynamics as in Eq. \ref{dyneta} in the presence of environmental noise introduced in Eq.\ \ref{multnoise}. The adaptation rule responds only to the deterministic component of the growth rate, i.e. environmental noise affects adaptation indirectly through changes in resource availability. Interestingly, environmental noise becomes beneficial at strong adaptation rates, as illustrated in Fig.\ \ref{adapt}(e). Figure\ \ref{adapt}(e) shows that an optimal intermediate adaptation rate exists for weak noise. As noise increases, higher adaptation rates are favored, while for very strong noise, no adaptation rate is beneficial. This highlights a beneficial role of environmental variability in the presence of spatial adaptation.


\section{Discussion}

In this work, we have presented a simple theoretical model for spatially structured microbial communities. Extending the consumer-resource model, we coarse-grained the effects of space on community dynamics into one parameter, $\theta$, which determines the degree to which resource competition is local versus global. This simple model accounts for the fact that space sets the scale over which species compete for nutrients. In heavily structured populations, where the community is fully separated into spatial niches ($\theta \rightarrow 1$), competition for nutrients is mostly local and direct competition between species is reduced. In well-mixed populations ($\theta \rightarrow 0$), competition is global and the competitive exclusion principle applies, with the most competitive species dominating the population. Avoiding explicit modeling of space allows us to explore several important consequences of spatial structure on microbial communities. In particular, using our model, we showed that increasing localization promotes coexistence, buffers communities from environmental fluctuations, synergistically increases the benefit of poor nutrients, and creates nontrivial effects when spatial configuration is allowed to adapt.

Spatial structure promotes coexistence by reducing global competition between species. Moreover, coexistence depends on how permissive the environment is to nutrient exchange between the global environment and local niches, with an optimal exchange rate maximizing coexistence. The same decoupling also decreases the propagation of perturbations through the community, making it less susceptible to perturbations. This becomes particularly important in fluctuating environments, where spatial separation can shield different parts of the community from environmental fluctuations and reduce the effect of random growth advantages that can allow some species to dominate the population. These results are consistent with previous work showing that spatial structure can promote coexistence \cite{lobanov2023spatial,kim2008defined}. Another effect of spatial structure is that it allows communities to benefit more from weakly preferred nutrients. With increasing spatial separation, even weakly preferred nutrients can contribute to sustaining additional local niches, allowing more species to persist.

Allowing species to adapt their spatial configuration introduces a new dynamical effect, where the rate of adaptation itself strongly affects coexistence. We find that intermediate adaptation rates promote coexistence, while very slow adaptation does not allow species to respond before competitive exclusion occurs, and very fast adaptation drives species toward extreme spatial configurations (either full isolation or full overlap) that reduce coexistence by allowing more competitive species to remain near $\theta_i=0$. Interestingly, environmental fluctuations change this picture, as stronger adaptation can become beneficial at low to intermediate noise strengths, while very strong noise is harmful regardless of the adaptation rate. Therefore, the effect of spatial adaptation on coexistence depends strongly on both the adaptation rate and the variability of the environment. Experimentally testing this interaction between environmental fluctuations and spatial adaptation is an interesting avenue for future investigation.

Our model presented here prioritized simplicity and focused on one effect of space, that it sets the scale of resource competition in the community, to allow further exploration of properties of microbial communities. However, in doing so, we neglected more complicated spatial dynamics that may affect microbial communities, such as space geometry, motility, within-niche interactions, and heterogeneity between local niches, as well as non-spatial processes that interact with spatial structure, such as cross-feeding. These added complications are beyond the scope of this work, but provide interesting avenues for future research that can build upon the current framework to investigate their interaction with space. Finally, another interesting avenue for future research is to utilize the simplicity of this model to study the effect of spatial structure on the evolution of microbial populations.

\section*{Acknowledgments}
We thank Obinna A. Ukogu for helpful feedback on the manuscript. This work was supported by the National Institutes of Health (Grant No. R35GM156451), the National Science Foundation (Grant No. DMS-2245816), and the Tarbutton Postdoctoral Fellowship from Emory University. This research was supported in part by the University of Pittsburgh Center for Research Computing and Data, RRID:SCR\_022735, through the resources provided. Specifically, this work used the H2P cluster, which is supported by NSF award number OAC-2117681, and the HTC cluster, which is supported by NIH award number S10OD028483. Code associated with this manuscript was written, revised, and optimized with assistance from OpenAI Codex (GPT-5.6 Sol).

\section*{Data availability}
The code and simulation data that support the findings of this article are openly available \cite{data}.

\bibliography{refs}

@misc{supp,
  note={See supplemental material for additional derivations and simulations.}
  }

@misc{data,
  note={\url{https://github.com/motasemelgamel/spatial-structure}}
}

@misc{paramchoice,
  note = {Unless otherwise stated or varied in the figure, the default simulation settings used throughout the paper were as follows: the community was initialized with $50$ species and $1$ nutrient. Model parameters were chosen as $\beta=1$, $\gamma=0.01$ concentration/cell, $k_i=1$, $\alpha_i$ was sampled uniformly between $1$ and $5$ for different species, and $R_0=5$ nutrient/volume. An abundance threshold of $N_i=10^{-3}$ was applied, below this threshold a species was considered extinct. All species had the same value of $\theta_i=\theta$.}
}

@article{goyal2025universal,
  title={Universal niche geometry governs the response of ecosystems to environmental perturbations},
  author={Goyal, Akshit and Rocks, Jason W and Mehta, Pankaj},
  journal={PRX life},
  volume={3},
  number={1},
  pages={013010},
  year={2025},
  publisher={APS}
}

@article{sayin2026bacterial,
  title={Bacterial population dynamics during colonization of solid tumors},
  author={Sayin, Serkan and ElGamel, Motasem and Rosener, Brittany and Brehm, Michael and Mugler, Andrew and Mitchell, Amir},
  journal={Molecular Systems Biology},
  volume={22},
  number={3},
  pages={412--434},
  year={2026},
  publisher={Springer}
}

@article{bar2018biomass,
  title={The biomass distribution on Earth},
  author={Bar-On, Yinon M and Phillips, Rob and Milo, Ron},
  journal={Proceedings of the National Academy of Sciences},
  volume={115},
  number={25},
  pages={6506--6511},
  year={2018},
  publisher={National Academy of Sciences}
}

@article{widder2016challenges,
  title={Challenges in microbial ecology: building predictive understanding of community function and dynamics},
  author={Widder, Stefanie and Allen, Rosalind J and Pfeiffer, Thomas and Curtis, Thomas P and Wiuf, Carsten and Sloan, William T and Cordero, Otto X and Brown, Sam P and Momeni, Babak and Shou, Wenying and others},
  journal={The ISME journal},
  volume={10},
  number={11},
  pages={2557--2568},
  year={2016},
  publisher={Oxford University Press}
}

@article{van2022ecological,
  title={Ecological modelling approaches for predicting emergent properties in microbial communities},
  author={van den Berg, Naomi Iris and Machado, Daniel and Santos, Sophia and Rocha, Isabel and Chac{\'o}n, Jeremy and Harcombe, William and Mitri, Sara and Patil, Kiran R},
  journal={Nature ecology \& evolution},
  volume={6},
  number={7},
  pages={855--865},
  year={2022},
  publisher={Nature Publishing Group UK London}
}

@article{warrier2026interplay,
  title={Interplay of Spatial Structure and Interactions in Microbial Communities},
  author={Warrier, Vaishnavi and Chen, Yilin and Rappaport, Ethan and Haruta, Shin and Youk, Hyun and Momeni, Babak},
  journal={Environmental Microbiology},
  volume={28},
  number={3},
  pages={e70262},
  year={2026},
  publisher={Wiley Online Library}
}

@article{weiner2019spatial,
  title={Spatial ecology of territorial populations},
  author={Weiner, Benjamin G and Posfai, Anna and Wingreen, Ned S},
  journal={Proceedings of the National Academy of Sciences},
  volume={116},
  number={36},
  pages={17874--17879},
  year={2019},
  publisher={National Academy of Sciences}
}

@article{kim2008defined,
  title={Defined spatial structure stabilizes a synthetic multispecies bacterial community},
  author={Kim, Hyun Jung and Boedicker, James Q and Choi, Jang Wook and Ismagilov, Rustem F},
  journal={Proceedings of the National Academy of Sciences},
  volume={105},
  number={47},
  pages={18188--18193},
  year={2008},
  publisher={National Academy of Sciences}
}

@article{schlomann2021gut,
  title={Gut bacterial aggregates as living gels},
  author={Schlomann, Brandon H and Parthasarathy, Raghuveer},
  journal={Elife},
  volume={10},
  pages={e71105},
  year={2021},
  publisher={eLife Sciences Publications, Ltd}
}

@article{henderson2025disentangling,
  title={Disentangling the feedback loops driving spatial patterning in microbial communities},
  author={Henderson, Alyssa and Del Panta, Alessia and Schubert, Olga T and Mitri, Sara and van Vliet, Simon},
  journal={npj Biofilms and Microbiomes},
  volume={11},
  number={1},
  pages={32},
  year={2025},
  publisher={Nature Publishing Group UK London}
}

@article{shi2020highly,
  title={Highly multiplexed spatial mapping of microbial communities},
  author={Shi, Hao and Shi, Qiaojuan and Grodner, Benjamin and Lenz, Joan Sesing and Zipfel, Warren R and Brito, Ilana Lauren and De Vlaminck, Iwijn},
  journal={Nature},
  volume={588},
  number={7839},
  pages={676--681},
  year={2020},
  publisher={Nature Publishing Group UK London}
}

@article{raynaud2014spatial,
  title={Spatial ecology of bacteria at the microscale in soil},
  author={Raynaud, Xavier and Nunan, Naoise},
  journal={PloS one},
  volume={9},
  number={1},
  pages={e87217},
  year={2014},
  publisher={Public Library of Science San Francisco, USA}
}

@article{yanni2019drivers,
  title={Drivers of spatial structure in social microbial communities},
  author={Yanni, David and M{\'a}rquez-Zacar{\'\i}as, Pedro and Yunker, Peter J and Ratcliff, William C},
  journal={Current biology},
  volume={29},
  number={11},
  pages={R545--R550},
  year={2019},
  publisher={Elsevier}
}

@article{momeni2013strong,
  title={Strong inter-population cooperation leads to partner intermixing in microbial communities},
  author={Momeni, Babak and Brileya, Kristen A and Fields, Matthew W and Shou, Wenying},
  journal={elife},
  volume={2},
  pages={e00230},
  year={2013},
  publisher={eLife Sciences Publications, Ltd}
}

@article{backer2026spatial,
  title={Spatial structure: shaping the ecology and evolution of microbial communities},
  author={B{\"a}cker, Marcel and Doekes, Hilje M and Garza, Daniel R and Meijer, Jeroen and van Vliet, Simon and Allen, Rosalind J and Hogeweg, Paulien and Dutilh, Bas E and van Dijk, Bram},
  journal={FEMS microbiology reviews},
  volume={50},
  pages={fuaf067},
  year={2026},
  publisher={Oxford University Press}
}

@article{lobanov2023spatial,
  title={Spatial structure favors microbial coexistence except when slower mediator diffusion weakens interactions},
  author={Lobanov, Alexander and Dyckman, Samantha and Kurkjian, Helen and Momeni, Babak},
  journal={Elife},
  volume={12},
  pages={e82504},
  year={2023},
  publisher={eLife Sciences Publications, Ltd}
}

@article{denk2025spatial,
  title={Spatial population dynamics of bacterial colonies with social antibiotic resistance},
  author={Denk-Lobnig, Marlis K and Wood, Kevin B},
  journal={Proceedings of the National Academy of Sciences},
  volume={122},
  number={7},
  pages={e2417065122},
  year={2025},
  publisher={National Academy of Sciences}
}

@article{al2026spatiotemporal,
  title={Spatiotemporal noise stabilizes unbounded diversity in strongly competitive communities},
  author={Al-Hiyasat, Amer and Swartz, Daniel W and Gore, Jeff and Kardar, Mehran},
  journal={Proceedings of the National Academy of Sciences},
  volume={123},
  number={41},
  pages={e2619598123},
  year={2026},
  publisher={National Academy of Sciences}
}

@article{garcia2024interactions,
  title={Interactions and migration rescuing ecological diversity},
  author={Garcia Lorenzana, Giulia and Altieri, Ada and Biroli, Giulio},
  journal={PRX Life},
  volume={2},
  number={1},
  pages={013014},
  year={2024},
  publisher={APS}
}

@article{cui2024houches,
  title={Les houches lectures on community ecology: From niche theory to statistical mechanics},
  author={Cui, Wenping and Marsland III, Robert and Mehta, Pankaj},
  journal={ArXiv},
  pages={arXiv--2403},
  year={2024}
}

@article{chesson1990macarthur,
  title={MacArthur's consumer-resource model},
  author={Chesson, Peter},
  journal={Theoretical Population Biology},
  volume={37},
  number={1},
  pages={26--38},
  year={1990},
  publisher={Elsevier}
}

@article{bunin2017ecological,
  title={Ecological communities with Lotka-Volterra dynamics},
  author={Bunin, Guy},
  journal={Physical Review E},
  volume={95},
  number={4},
  pages={042414},
  year={2017},
  publisher={APS}
}

@article{dal2020short,
  title={Short-range interactions govern the dynamics and functions of microbial communities},
  author={Dal Co, Alma and van Vliet, Simon and Kiviet, Daniel Johannes and Schlegel, Susan and Ackermann, Martin},
  journal={Nature ecology \& evolution},
  volume={4},
  number={3},
  pages={366--375},
  year={2020},
  publisher={Nature Publishing Group UK London}
}

@article{pickett1995landscape,
  title={Landscape ecology: spatial heterogeneity in ecological systems},
  author={Pickett, Steward TA and Cadenasso, Mary L},
  journal={Science},
  volume={269},
  number={5222},
  pages={331--334},
  year={1995},
  publisher={American Association for the Advancement of Science}
}

@article{hardin1960competitive,
  title={The competitive exclusion principle: an idea that took a century to be born has implications in ecology, economics, and genetics.},
  author={Hardin, Garrett},
  journal={science},
  volume={131},
  number={3409},
  pages={1292--1297},
  year={1960},
  publisher={American Association for the Advancement of Science}
}

@article{cui2020effect,
  title={Effect of resource dynamics on species packing in diverse ecosystems},
  author={Cui, Wenping and Marsland III, Robert and Mehta, Pankaj},
  journal={Physical review letters},
  volume={125},
  number={4},
  pages={048101},
  year={2020},
  publisher={APS}
}

@article{long2005aquatic,
  title={Aquatic habitats and diffusion constraints affecting microbial coexistence in unsaturated porous media},
  author={Long, Tao and Or, Dani},
  journal={Water Resources Research},
  volume={41},
  number={8},
  year={2005},
  publisher={Wiley Online Library}
}

@article{mattei2024exploring,
  title={Exploring spatial segregation induced by competition avoidance as driving mechanism for emergent coexistence in microbial communities},
  author={Mattei, Mattia and Arenas, Alex},
  journal={Physical Review E},
  volume={110},
  number={1},
  pages={014404},
  year={2024},
  publisher={APS}
}

@article{xiao2026environmental,
  title={Environmental stochasticity can account for patterns of within-host respiratory virus evolution},
  author={Xiao, Wenfei Fiona and Farjo, Mireille and Lowen, Anice and Koelle, Katia},
  journal={bioRxiv},
  pages={2026--05},
  year={2026},
  publisher={Cold Spring Harbor Laboratory}
}

@article{mancuso2021environmental,
  title={Environmental fluctuations reshape an unexpected diversity-disturbance relationship in a microbial community},
  author={Mancuso, Christopher P and Lee, Hyunseok and Abreu, Clare I and Gore, Jeff and Khalil, Ahmad S},
  journal={Elife},
  volume={10},
  pages={e67175},
  year={2021},
  publisher={eLife Sciences Publications, Ltd}
}

@article{keegstra2022ecological,
  title={The ecological roles of bacterial chemotaxis},
  author={Keegstra, Johannes M and Carrara, Francesco and Stocker, Roman},
  journal={Nature Reviews Microbiology},
  volume={20},
  number={8},
  pages={491--504},
  year={2022},
  publisher={Nature Publishing Group UK London}
}

@article{henderson2026fluctuating,
  title={Fluctuating environments are sufficient to drive substantial variability in species abundance across locations},
  author={Henderson, James FD and Tiffeau-Mayer, Andreas},
  journal={arXiv preprint arXiv:2603.02403},
  year={2026}
}

@article{mattei2026nonlinear,
  title={Nonlinear competition avoidance favors coexistence in microbial populations},
  author={Mattei, Mattia and Soriano-Pa{\~n}os, David and Arenas, Alex},
  journal={arXiv preprint arXiv:2601.13947},
  year={2026}
}

@article{mark2016biogeography,
  title={Biogeography of a human oral microbiome at the micron scale},
  author={Mark Welch, Jessica L and Rossetti, Blair J and Rieken, Christopher W and Dewhirst, Floyd E and Borisy, Gary G},
  journal={Proceedings of the National Academy of Sciences},
  volume={113},
  number={6},
  pages={E791--E800},
  year={2016},
  publisher={National Academy of Sciences}
}

\clearpage
\onecolumngrid
\beginsupplement

\begin{center}
{\bf SUPPLEMENTAL MATERIAL}
\end{center}

\section{Model}

The model before rescaling is given by
\begin{gather}
    \frac{dN_i}{dt}=[(1-\theta_i)r_i R + \theta_i r_i T_i -m_i] N_i, \label{supp:spatialCRmdlNor}\\
    \frac{dR}{dt}=-\gamma R \sum_{i=1}^S (1-\theta_i) r_i N_i -D \sum_{i=1}^S (R-T_i) + D_0(R_0 - R), \label{supp:spatialCRmdlRor}\\
    \frac{dT_i}{dt}=D (R-T_i) -\gamma \theta_i r_i T_i N_i. \label{supp:spatialCRmdlTor}
\end{gather}

\section{Coexistence}

We start with the rescaled model introduced in the main text. The model is given by

\begin{gather}
    \frac{dN_i}{d\tau}=[(1-\theta_i)\alpha_i R + \theta_i \alpha_i T_i -k_i] N_i, \label{supp:spatialCRmdlN}\\
    \frac{dR}{d\tau}=-\gamma R \sum_{i=1}^S (1-\theta_i) \alpha_i N_i -\beta \sum_{i=1}^S (R-T_i) + (R_0 - R), \label{supp:spatialCRmdlR}\\
    \frac{dT_i}{d\tau}=\beta (R-T_i) -\gamma \theta_i \alpha_i T_i N_i. \label{supp:spatialCRmdlT}
\end{gather}
For tractability, we study the simple case of two species competing for one nutrient in a spatially structured environment. We assume that both species have the same degree of separation in the environment, $\theta_1=\theta_2=\theta$. The full model describing species dynamics is given by
\begin{gather}
    \frac{dN_1}{d\tau}=[(1-\theta)\alpha_1 R + \theta \alpha_1 T_1 -k_1] N_1, \label{species1}\\
    \frac{dN_2}{d\tau}=[(1-\theta)\alpha_2 R + \theta \alpha_2 T_2 -k_2] N_2, \label{species2}\\
    \frac{dR}{d\tau}=-\gamma R (1-\theta) (\alpha_1 N_1+\alpha_2 N_2) -\beta (2R-T_1-T_2) + (R_0 - R), \label{Rdyn}\\
    \frac{dT_1}{d\tau}=\beta (R-T_1) -\gamma \theta \alpha_1 T_1 N_1, \label{T1dyn}\\
    \frac{dT_2}{d\tau}=\beta (R-T_2) -\gamma \theta \alpha_2 T_2 N_2. \label{T2dyn}
\end{gather}
Next, we study the well-mixed ($\theta=0$) and the fully separated limits ($\theta=1$). In the well-mixed limit at steady state, and assuming $N_1>0$ and $N_2>0$, we find, using eqs.\ \ref{species1} and \ref{species2} and $T_i=R$,
\begin{gather}
    \alpha_1 R -k_1=0, \label{species1ss}\\
    \alpha_2 R -k_2=0, \label{species2ss}\\
    -\gamma R (\alpha_1 N_1+\alpha_2 N_2) + (R_0 - R)=0 \label{Rdynss}
\end{gather}
Eqs.\ \ref{species1ss} and\ \ref{species2ss} can only be satisfied if $\frac{k_1}{\alpha_1}=\frac{k_2}{\alpha_2}$, i.e. only neutral coexistence is possible in the well-mixed case and stable coexistence is generally not possible. We now focus on the fully separated case ($\theta=1$) at steady state. Assuming $N_1>0$ and $N_2>0$, the model becomes
\begin{gather}
    \alpha_1 T_1 -k_1=0, \label{species1ss2}\\
    \alpha_2 T_2 -k_2=0, \label{species2ss2}\\
    -\beta (2R-T_1-T_2) + (R_0 - R)=0, \label{Rdynss2}\\
    \beta (R-T_1) -\gamma \alpha_1 T_1 N_1=0, \label{T1dynss2}\\
    \beta (R-T_2) -\gamma \alpha_2 T_2 N_2=0. \label{T2dynss2}
\end{gather}
Solving this system of equations yields

\begin{gather}
    T_1=\frac{k_1}{\alpha_1}, \label{T1ss}\\
    T_2=\frac{k_2}{\alpha_2}, \label{T2ss}\\
    R=\frac{1}{1+2\beta}[R_0+\beta(\frac{k_1}{\alpha_1}+\frac{k_2}{\alpha_2})], \label{Rss}\\
    N_1=\frac{\beta}{\gamma k_1 (1+2 \beta)} [R_0 - (1+\beta) \frac{k_1}{\alpha_1} + \beta \frac{k_2}{\alpha_2}], \label{N1ss}\\
    N_2=\frac{\beta}{\gamma k_2 (1+2 \beta)} [R_0 - (1+\beta) \frac{k_2}{\alpha_2} + \beta \frac{k_1}{\alpha_1}]. \label{N2ss}
\end{gather}
The condition for coexistence then becomes $R_0>(1+\beta) k_1/\alpha_1 - \beta k_2/\alpha_2$ and $R_0>(1+\beta) k_2/\alpha_2 - \beta k_1/\alpha_1$, which can be simultaneously satisfied for sufficiently large $R_0$. In the limit $\beta \rightarrow \infty$, the condition cannot be simultaneously satisfied and the species with larger ratio, $k_i / \alpha_i$, will go extinct. Intuitively, when the exchange rate is large, the community returns to the well-mixed case, spatial separation is no longer beneficial, and competitive exclusion applies. In the limit $\beta\rightarrow0$, both $N_1\rightarrow0$ and $N_2\rightarrow0$, due to starvation from insufficient nutrient diffusion into their local niches.

\section{Synergy with additional nutrients}
Disclosure: derivations in this section used the assistance of GPT-6.1 Sol.

\begin{figure}[H]
\centering
\includegraphics[width=1\linewidth]{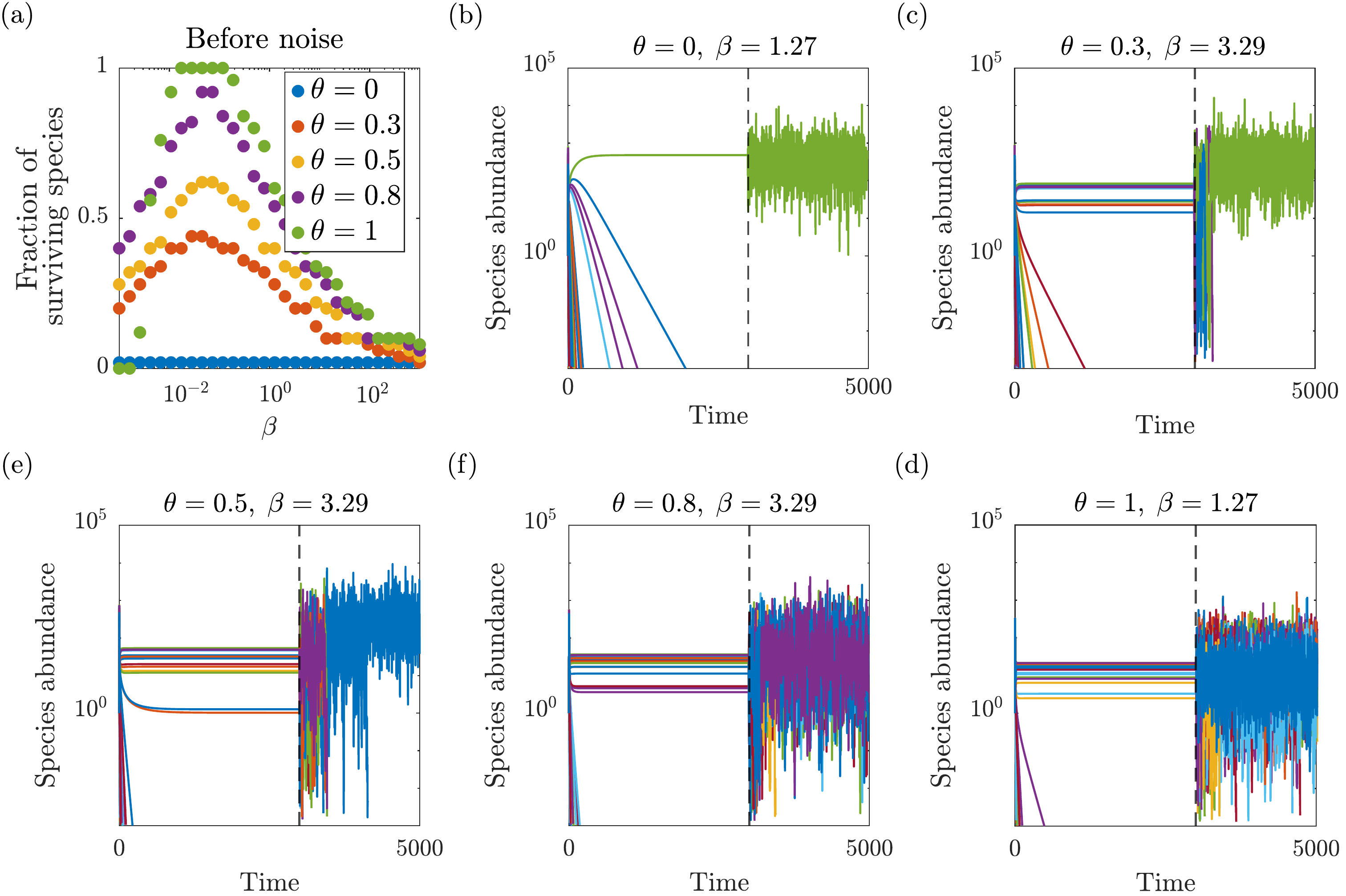}
\caption{(a) Fraction of surviving species vs. $\beta$, shown for different values of $\theta$ before adding noise. (b)-(d) Species abundance dynamics before and after adding environmental noise. The black dashed vertical line indicates the time at which environmental noise was introduced.}
\label{Sfig1:noise}
\end{figure}

Here, we derive a lower bound on the number of surviving species in the presence of poor nutrients and spatial structure. We use the generalized model introduced in the main text to account for an arbitrary number of nutrients and species. We state the model here again:
\begin{gather}
    \frac{dN_i}{d\tau}=[(1-\theta_i) \sum_{j=1}^n \delta_{ij} R_j+\theta_i \sum_{j=1}^n \delta_{ij} T_{ij}   -k_i] N_i, \label{spatialmultinutrientNsupp}\\
     \frac{dR_j}{d\tau}=-\gamma R_j \sum_{i=1}^S (1-\theta_i) \delta_{ij} N_i -\beta \sum_{i=1}^S (R_j-T_{ij})+ (R_{j0} - R_j), \label{spatialmultinutrientRsupp}\\
    \frac{dT_{ij}}{d\tau}=\beta (R_j-T_{ij}) -\gamma \theta_i \delta_{ij} T_{ij} N_i, \label{spatialmultinutrientTsupp}
\end{gather}
where $R_j$ denotes the global concentration of nutrient $j$, $T_{ij}$ is the concentration of nutrient $j$ in niche $i$, $\delta_{ij}$ is the preference of species $i$ for nutrient $j$, $j$ is an index over nutrients, and $n$ is the total number of nutrients. Here, $S$ is the initial number of species. It is assumed that all nutrients diffuse into the local niches at the same rate $\beta$. Similarly to the main text, we consider only one resource for which all species have a strong preference (rich nutrient). All other added resources are weakly preferred by the species (poor nutrients). We refer to the species' preferences for the rich nutrient as $\alpha$, as used previously, while $\delta$ will refer to the species' preferences for poor nutrients. Specializing to $\theta=0$, using $T_{ij}=R_j$, the model becomes
\begin{gather}
    \frac{dN_i}{d\tau}=[ \sum_{j=1}^n \delta_{ij} R_j  -k_i] N_i, \label{spatialmultinutrientNsupp1}\\
     \frac{dR_j}{d\tau}=-\gamma R_j \sum_{i=1}^S \delta_{ij} N_i + (R_{j0} - R_j). \label{spatialmultinutrientRsupp1}
\end{gather}
We consider the simplified case in which all species have the same preferences for poor nutrients ($R_j$), $\delta_{ij}=\delta$ for $j\geq2$, and different preferences for one rich nutrient ($R_1$), $\delta_{i1}=\alpha_i$. We also set $R_{j0}=R_0$ and $k_i=k$. These equations become
\begin{gather}
    \frac{dN_i}{d\tau}=[\alpha_i R_1 + \delta \sum_{j=2}^n R_j  -k] N_i, \label{spatialmultinutrientNsupp2}\\
     \frac{dR_j}{d\tau}=-\gamma R_j \delta \sum_{i=1}^S N_i + (R_{0} - R_j), \qquad j=2,\ldots,n, \label{spatialmultinutrientRsupp2}\\
     \frac{dR_1}{d\tau}=-\gamma R_1 \sum_{i=1}^S \alpha_i N_i + (R_{0} - R_1). \label{spatialmultinutrientRsupp3}
\end{gather}
At steady state, assuming $N_i>0$ and $N_l>0$, we consider eq.\ \ref{spatialmultinutrientNsupp2} for two species $i$ and $l$:
\begin{gather}
    \alpha_i R_1 + \delta \sum_{j=2}^n R_j  -k=0, \label{spatialmultinutrientNsuppi}\\
    \alpha_l R_1 + \delta \sum_{j=2}^n R_j  -k=0. \label{spatialmultinutrientNsuppl}
\end{gather}
Subtracting eqs.\ \ref{spatialmultinutrientNsuppi} and \ref{spatialmultinutrientNsuppl}, we find that, with $R_1>0$, coexistence requires $\alpha_i=\alpha_l$ regardless of the number of poor nutrients present, $n-1$. Next, we consider the $\theta=1$ case. The model reads
\begin{gather}
    \frac{dN_i}{d\tau}=[\alpha_i T_{i1} + \delta \sum_{j=2}^n T_{ij}   -k] N_i, \label{spatialmultinutrientNsupp}\\
     \frac{dR_j}{d\tau}=-\beta \sum_{i=1}^S (R_j-T_{ij})+ (R_{0} - R_j), \label{spatialmultinutrientRsupp}\\
    \frac{dT_{ij}}{d\tau}=\beta (R_j-T_{ij}) -\gamma \delta_{ij} T_{ij} N_i. \label{spatialmultinutrientTsupp}
\end{gather}
At steady state, an extinct species has $N_i=0$, so its local nutrient concentration equals the global concentration, $T_{ij}=R_j$. Therefore, the term $\sum_{i=1}^S T_{ij}$ in eq.\ \ref{spatialmultinutrientRsupp} can be written as $\sum_{i=1}^{\tilde{S}} T_{ij} + (S-\tilde{S})R_j$. Hence, eq.\ \ref{spatialmultinutrientRsupp} at steady state gives
\begin{gather}
    R_j=\frac{R_0 + \beta \sum_{i=1}^{\tilde{S}} T_{ij} }{1+\beta \tilde{S}}, \label{Rjs}
\end{gather}
where $\tilde{S}$ is the number of surviving species, and the survivors have been relabeled $i=1,\ldots,\tilde S$. The following inequality holds for both rich and poor nutrients:
\begin{gather}
    R_j=\frac{R_0 + \beta \sum_{i=1}^{\tilde{S}} T_{ij} }{1+\beta \tilde{S}} \ge \frac{R_0 }{1+\beta \tilde{S}}. \label{Rjs2}
\end{gather}
If a species is extinct at equilibrium, its growth rate when introduced at a very low abundance must be non-positive to stay extinct and must satisfy
\begin{gather}
\alpha_i R_1 + (n-1) \delta R_j \le k, \label{ineq}
\end{gather}
where we used $T_{i1}=R_1$ and $T_{ij}=R_j$ for an extinct species. All poor nutrients have the same global steady-state concentration $R_j$ under the common-preference assumptions above. Both inequalities, \ref{Rjs2} and \ref{ineq}, when combined give a lower bound on the number of surviving species
\begin{gather}
\tilde{S}\ge \frac{R_0(\alpha_i + (n-1) \delta)-k}{\beta k}. \label{survive}
\end{gather}
This shows that the lower bound on surviving species increases with $n-1$, the number of poor nutrients, allowing more species to coexist. It does not, by itself, prove that the exact survivor count increases with $n$. This effect emerged for $\theta=1$ and was not possible for $\theta=0$, illustrated in figs.\ \ref{Sfig1:noise} and\ \ref{hetenv}. Intermediate $\theta$ values also allow for greater coexistence but an analysis of the general case would be significantly more mathematically complicated.

\section{Susceptibility}

We analyze the susceptibility of two species to parameter perturbations following the analysis of ref.\ \cite{goyal2025universal}. The model for two species is given by eqs.\ \ref{species1}-\ref{T2dyn}. Next, we introduce a perturbation in the mortality rates, $k_1\rightarrow k_1 + \delta k_1$ and $k_2\rightarrow k_2 + \delta k_2$. As a result, the system will move to a new steady state given by $N_1 \rightarrow N_1 + \delta N_1$, $N_2 \rightarrow N_2 + \delta N_2$, $T_1 \rightarrow T_1 + \delta T_1$, $T_2 \rightarrow T_2 + \delta T_2$, and $R \rightarrow R + \delta R$. The susceptibilities are then defined by the ratio of the resulting change in steady state (the response) to the perturbation, e.g. $\delta N_1/\delta k_1$, and similarly for other variables. Here, we focus on the susceptibilities of the abundances to mortality-rate perturbations, and define $\chi_{ij}=\delta N_i/\delta k_j$, with all other parameters held fixed. This linear response assumes that both species remain present. We substitute the perturbed steady-state values into eqs.\ \ref{species1}-\ref{T2dyn} which gives

 \begin{gather}
(1-\theta)\alpha_1 (R+\delta R) + \theta \alpha_1 (T_1+\delta T_1) -(k_1+\delta k_1)=0, \label{species1perturb}\\
(1-\theta)\alpha_2 (R+\delta R) + \theta \alpha_2 (T_2+\delta T_2) -(k_2+\delta k_2)=0, \label{species2perturb}\\
 -\gamma (R+\delta R)  (1-\theta) (\alpha_1 (N_1+\delta N_1)+\alpha_2 (N_2+\delta N_2)) -\beta (2R+2 \delta R-T_1-\delta T_1-T_2-\delta T_2) + (R_0 - R-\delta R)=0, \label{Rdynperturb}\\
\beta (R+\delta R-T_1-\delta T_1) -\gamma \theta \alpha_1 (T_1+\delta T_1)(N_1+\delta N_1)=0, \label{T1dynperturb}\\
\beta (R+\delta R-T_2-\delta T_2) -\gamma \theta \alpha_2 (T_2+\delta T_2)(N_2+\delta N_2)=0. \label{T2dynperturb}
\end{gather}
The perturbations are assumed to be small relative to the corresponding steady-state values. Subtracting the unperturbed steady-state equations eliminates the zeroth-order terms. We retain only first-order terms and obtain the following system of equations:
 \begin{gather}
\alpha_1(\theta \delta T_1 + (1-\theta) \delta R)-\delta k_1=0, \label{species1delta}\\
\alpha_2(\theta \delta T_2 + (1-\theta) \delta R)-\delta k_2=0, \label{species2delta}\\
 -\gamma (1-\theta)(\alpha_1 N_1 \delta R + \alpha_2 N_2 \delta R + \alpha_1 R \delta N_1 + \alpha_2 R \delta N_2) -\beta (2 \delta R-\delta T_1-\delta T_2) -\delta R=0, \label{Rdyndelta}\\
\beta (\delta R-\delta T_1) -\gamma \theta \alpha_1 (T_1 \delta N_1 +N_1 \delta T_1)=0, \label{T1dyndelta}\\
\beta (\delta R-\delta T_2) -\gamma \theta \alpha_2 (T_2 \delta N_2+N_2 \delta T_2)=0. \label{T2dyndelta}
\end{gather}
Solving these equations gives $\delta N_i=\chi_{ii} \delta k_i +\chi_{ij} \delta k_j$, where $j\neq i$ and the coefficients $\chi_{ij}$ are the susceptibilities of the community to perturbations in the mortality rates. We solve this system computationally \cite{data} and plot the susceptibility of species 1 to a perturbation in its own mortality rate in fig.\ \ref{susc}(a).

\newpage
\section{supplementary figures}

\begin{figure}[H]
\centering
\includegraphics[width=1\linewidth]{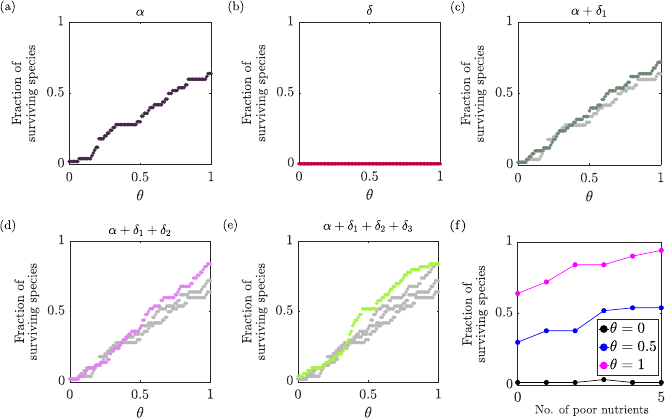}
\caption{Fraction of surviving species as a function of $\theta$ in the presence of only one nutrient for which all species have (a) a strong (rich nutrient) (b) or weak (poor nutrient) preference.  No species survive in the presence of only the poor nutrient. (c)-(f) Sequentially adding poor nutrients to the environment increases the number of surviving species in the presence of the rich nutrient compared to the original fraction of surviving species in the presence of only the rich nutrient. In the well-mixed limit ($\theta=0$), the presence of both nutrients does not appreciably increase survival, this synergistic effect only appears in the presence of spatial structure. Species preferences for the poor nutrient were sampled uniformly between $\delta=0.05$ and $\delta=0.1$.}
\label{Sfig2:hetenvpoor}
\end{figure}

\begin{figure}[H]
\centering
\includegraphics[width=0.6\linewidth]{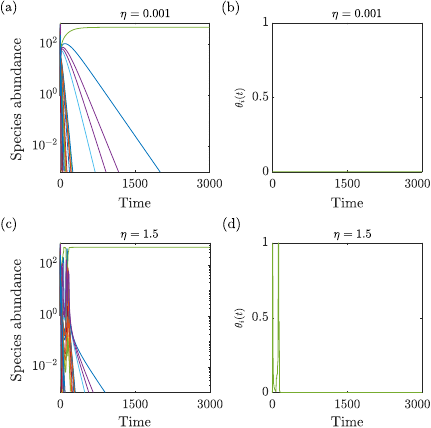}
\caption{(a)-(b) Abundance and $\theta$ dynamics for a slow adaptation rate, $\eta=0.001$. (c)-(d) Abundance and $\theta$ dynamics for a fast adaptation rate, $\eta=1.5$. All species were initialized with an abundance of one. Neither slow nor fast adaptation rates are beneficial for species survival, as explained in the main text.}
\label{Sfig3:adaptlowhigh}
\end{figure}

\begin{figure}[H]
\centering
\includegraphics[width=0.6\linewidth]{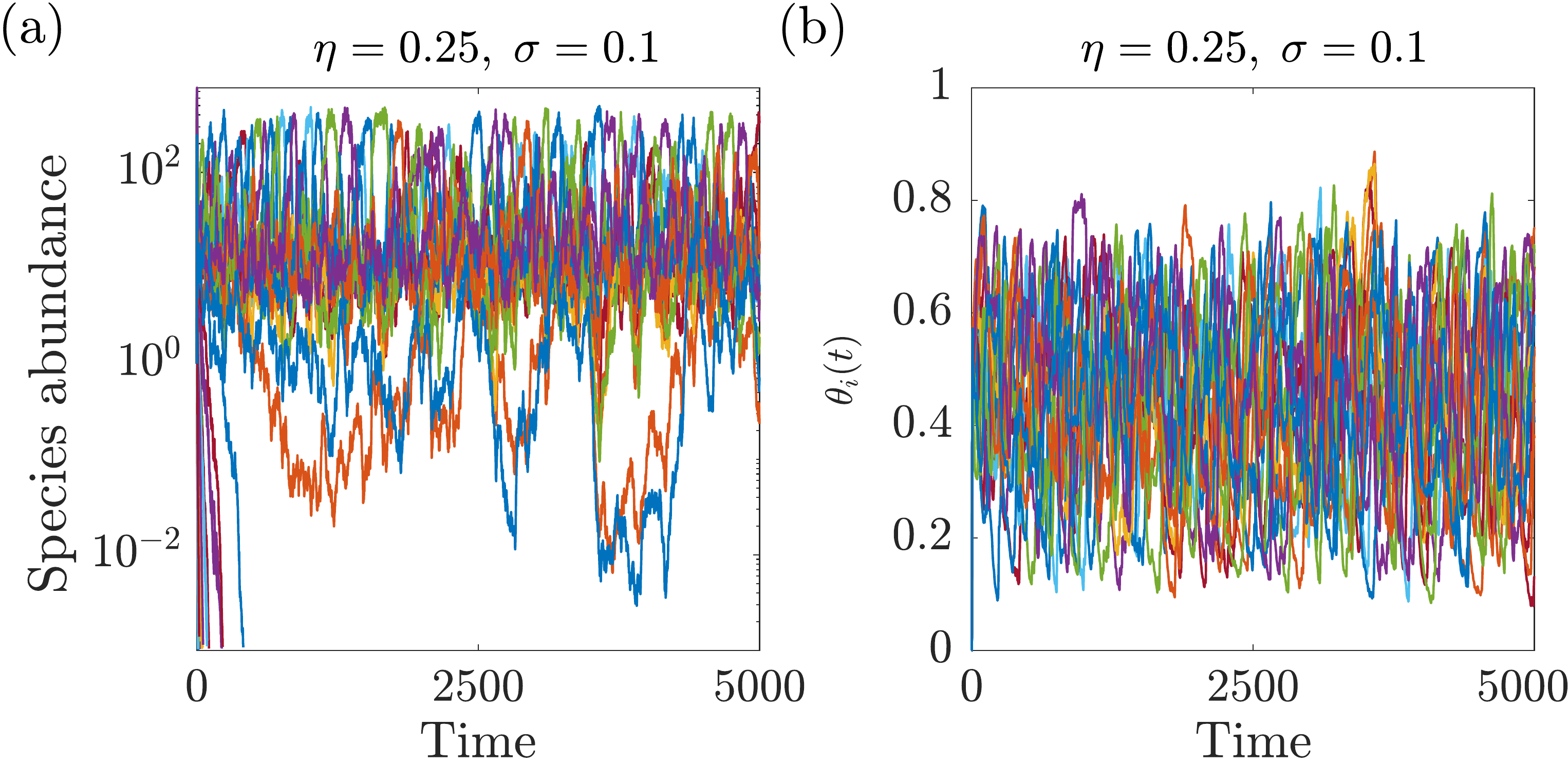}
\caption{(a)-(b) Abundance and $\theta$ dynamics for a moderate adaptation rate, $\eta=0.25$, and environmental noise with strength $\sigma=0.1$. All species were initialized with an abundance of one.}
\label{Sfig4:adaptwithnoise}
\end{figure}

\end{document}